\documentclass[prl,aps,twocolumn,groupedaddress,showpacs,floatfix,preprintnumbers]{revtex4}
\usepackage{amsmath,amssymb,multirow,epsfig,bm}

\newcommand{\etal}{{\it et al.,\;}}
\newcommand{\avg}[1]{\langle #1 \rangle}
\newcommand{\eqn}[1]{Eq.~(\ref{#1})}
\newcommand{\fig}[1]{Fig.~\ref{#1}}
\newcommand{\apriori}{\textit{a~priori} }              
\newcommand{\pictsize}{1.0}
\newcommand{\beq}{\begin{equation}}
\newcommand{\eeq}{\end{equation}}
\newcommand{\bea}{\begin{eqnarray}}
\newcommand{\eea}{\end{eqnarray}}
\newcommand{\pcut}{k^{}_{\text{max}}}  
 
\newcommand{\eF}{\varepsilon^{}_{F}}
\newcommand{\pF}{k_{F}}
\newcommand{\reff}{r^{}_{\textrm{eff}}}

\begin{document}
\preprint{NT@UW-13-12}

\title{The temperature evolution of the shear viscosity in a unitary Fermi gas}

\author{Gabriel Wlaz\l{}owski$^{1,2}$, Piotr Magierski$^{1,2}$, Aurel Bulgac$^{2}$ and Kenneth J. Roche$^{2,3}$}

\affiliation{$^1$Faculty of Physics, Warsaw University of Technology, Ulica Koszykowa 75, 00-662 Warsaw, Poland}
\affiliation{$^2$Department of Physics, University of Washington, Seattle, Washington 98195--1560, USA}
\affiliation{$^3$Pacific Northwest National Laboratory, Richland, Washington 99352, USA}

\begin{abstract}
We present an {\it ab initio} determination of the shear viscosity for the unitary Fermi gas based on finite temperature quantum Monte Carlo (QMC) calculations and the Kubo linear-response formalism. The results are confronted with the bound 
for the shear viscosity originating from hydrodynamic fluctuations. Assuming smoothness of the frequency dependent shear viscosity $\eta(\omega)$, we show that the bound is violated in the low temperature regime and 
the violation occurs simultaneously with the onset of the Cooper paring in the system. 
In order to preserve the hydrodynamic bound in QMC $\eta(\omega)$ has to possess a sharp structure located 
in the vicinity of zero frequency which is not resolved by an analytic continuation procedure.
\end{abstract}

\date{\today}

\pacs{03.75.Ss, 05.30.Fk, 05.60.Gg, 51.20.+d}

\maketitle

\section{Introduction}
The unitary Fermi gas~(UFG) is expected to be one of the most strongly correlated systems in nature as 
it saturates the unitarity bound for the $s$-wave 
cross section $\sigma(k) \leq 4\pi / k^2$, where $k$ is the relative wave vector of scattering particles.
The strong correlations are responsible for a multitude of interesting phenomena. The most surprising ones
include the existence of the pseudogap regime between the superfluid state and the normal state~\cite{stewart2010,gaebler2010,perali} and the nearly ideal hydrodynamic behavior~\cite{Turlapov,Cao1,Cao2}.
Therefore the properties of the UFG attract enormous attention from several communities including atomic physics, nuclear physics,
relativistic heavy-ion collisions, and high-$T_c$ superconductivity (see review papers \cite{reviews1,reviews2,SchaferTeaney,bcsbec}).

Experimentally the UFG has been realized with trapped fermionic atoms by means of Feshbach resonances~\cite{FirstExperiments} and currently represents one of the most controllable quantum systems.
However, a theoretical description poses a real challenge due to the lack of any 
small, dimensionless expansion parameter. Fortunately, for unpolarized UFG 
the genuine nonperturbative quantum Monte Carlo (QMC) techniques are free from the infamous fermionic sign problem, 
and can be used to produce reliable predictions, providing a solid benchmark for experimental as well as other 
theoretical studies. 
In particular, the energy of the ground state agrees with experimental 
results within $1\%$\cite{Carlsonetal} and the same accuracy was obtained for the equation of state for the system 
being in the normal state~\cite{VanHouckeetal}. For temperatures around the critical temperature the discrepancy between experiment and the simulations does not exceed $5\%$~\cite{Drutetal}. 
Other properties well established by QMC include the evolution of the pairing gap~\cite{CarlsonReddy,GezerlisCarlson,mag2009,mag2011}, the critical temperature for the superfluid-to-normal phase transition~\cite{Burovskietal,Burovskietal2,BDM},  and the calculation 
of transport coefficients of the UFG, namely, the shear viscosity~\cite{Wlazlowskietal} and the spin-diffusion coefficient~\cite{Wlazlowskietal2}.

Determination of transport coefficients is currently of great interest as these provide a verification of various 
hypothetical bounds having their roots directly in quantum mechanics. The best known example is a conjecture formulated 
by  Kovtun, Son, and Starinets (KSS) of the existence of a lower bound $\eta/s\geqslant\hbar/(4\pi k_{B})$  
on the ratio of the shear viscosity $\eta$ to the entropy density $s$, for all fluids~\cite{KSS}.
Due to  strong correlations which imply short mean free path,
the UFG appears as one of the best candidates for being a ``perfect fluid'' (besides the quark-gluon plasma created in heavy ion collisions), defined as the one with the lowest ratio $\eta/s$. Experimental measurements for trapped systems yield $\eta/s\leqslant 0.5\,\hbar/k^{}_{B}$~\cite{Turlapov,Cao1,Cao2,SchaferTeaney}, while the recent QMC simulations 
set the minimum value to be $\eta/s\approx 0.2\,\hbar/k^{}_{B}$~\cite{Wlazlowskietal}.
Still there are contradicting scenarios of how the shear viscosity behaves in the low temperature regime. 
Whereas some authors expect the pairing correlations to reduce the viscosity significantly and even 
to decrease it to zero at $T=0$~\cite{BruunSmith,Guoetal}, others favor the upturn originating from the phonon contribution~\cite{RupakSchafer}.

Recently it was shown by Chafin and Sch\"{a}fer that classical hydrodynamic fluctuations in a non-relativistic fluid lead
to the appearance of the minimum in the shear viscosity to density ratio $\eta/n$ as a function of the temperature~\cite{ChafinSchafer}. In the vicinity of the critical temperature the bound was established to be 
$\eta/n\gtrsim0.3\,\hbar$. This result was confirmed by Romatschke and Young~\cite{RomatschkeYoung}. While the experimental results for trapped systems are consistent with the hydrodynamic bound, the QMC results for uniform systems violate this bound as they predict values $\eta/n\lesssim0.2\,\hbar$ in the superfluid phase~\cite{Wlazlowskietal}. However, to extract the shear viscosity within QMC approach one has to perform an analytic continuation of the imaginary time correlator to real frequencies, which is a highly nontrivial procedure. It is known that this step represents an ill-posed numerical problem, where statistical fluctuations which are tied up with the QMC are greatly enhanced during the continuation process. 
Indeed,
it was pointed out in the supplement of~\cite{Wlazlowskietal}, and also in~\cite{RomatschkeYoung},
that the statistical accuracy of the quantum Monte Carlo signal
allows to produce the ratio $\eta/n$ consistent with the hydrodynamic
bound.

Here we present QMC results for a temperature evolution of the shear viscosity of higher accuracy than 
in our previous studies~\cite{Wlazlowskietal} and discuss the errors associated with the analytic continuation.
Under the assumption of the smoothness of frequency dependent shear viscosity $\eta(\omega)$,
we show that the new results violate the classical hydrodynamic bound \cite{ChafinSchafer,RomatschkeYoung} for $\eta/n$ in a low temperature regime, while the ratio $\eta/s$ stays above the KSS bound.  Moreover the hydrodynamic bound is violated for the temperatures at which
the Cooper pairs are formed in the system. We show that the agreement with hydrodynamic bound can be achieved only
if there exists a low frequency sharp peak in the $\eta(\omega)$ which is overlooked by analytic continuation procedure.

In order to determine the shear viscosity of the UFG we employ the QMC technique on the lattice, which provides numerical results with controllable accuracy, up to quantifiable systematic uncertainties (for details see Ref.~\cite{BDM}). These simulations are very similar to those of Ref.~\cite{Wlazlowskietal}
and we therefore only briefly describe the main stages of the computational process, 
focusing mainly on improvements. Henceforth we define units: $\hbar=m=k_{B}=1$.

\section{Method}

We performed simulations using three lattice sizes $N_x=8,10,12$ with corresponding average densities $n\backsimeq0.08, 0.04$ and $0.03$, respectively. For these lattices the systematic errors were estimated to be less than $10-15\%$ and are the most severe in the superfluid phase (for detailed discussion in the context of the transport coefficients see the Supplemental Material of Refs.~\cite{Wlazlowskietal,Wlazlowskietal2}). These errors are related mainly to corrections coming from the nonzero effective range $\reff$ and exclude the universal high momenta tail in the occupation probability by finite momentum cut-off $\pcut$. We obtain rather moderate values of $\pF\reff\backsimeq 0.54,\,0.43,\,0.39$ and $\pcut/\pF\backsimeq 2.4,\,3.0,\, 3.3$, respectively for lattices $N_x=8,\,10$ and $12$ [$\pF=(3\pi^{2}n)^{1/3}$ stands for Fermi momentum].
The statistical uncertainties are typically of the order of $1\%$ as an ensemble of $10^4$ uncorrelated samples were collected at each temperature. The fact that we have used three different lattices allows us to check the stability of the results as we approach the thermodynamic $V \to \infty$ and continuum $n \to 0$ limits, where $V=N_x^3$ (the lattice constant is set to unity). These new results were obtained with a new (parallelized) code and for densities significantly lower than those reported in Ref.~\cite{Wlazlowskietal}, and thus are closer to the desired limit $n \rightarrow 0$. The frequency dependent shear viscosity $\eta(\omega)$ is obtained from the imaginary-time (Euclidean) stress tensor-stress tensor correlator
\begin{equation}
 G_{\Pi}(\bm{q},\tau)=\dfrac{1}{V}
 \avg{\hat{\Pi}_{\bm{q}}^{(xy)}(\tau)\hat{\Pi}_{-\bm{q}}^{(xy)}(0)},
\label{eqn:GPi}
\end{equation} 
evaluated at zero momentum $\bm{q}=0$, by inversion of the relation
\begin{equation}
  G_{\Pi}(\bm{q}=0,\tau)=\dfrac{1}{\pi}
  \int_{0}^{\infty}\eta(\omega)\,
  \omega\dfrac{\cosh\left[ \omega(\tau-\beta/2)\right] }{\sinh\left( \omega\beta/2\right) }
  d\omega,
  \label{eqn:InversionProblem}
\end{equation}
where $\beta=1/T$ is the inverse of the temperature. Taking the limit of zero frequency we obtain the static shear viscosity $\eta\equiv\eta(\omega\to 0)$.
The average in \eqn{eqn:GPi} is performed over the grand canonical ensemble, 
$\hat{O}(\tau)=e^{\tau(\hat{H}-\mu\hat{N})}\hat{O}e^{-\tau(\hat{H}-\mu\hat{N})}$,
$\hat{H}$ is the Hamiltonian of the system, $\mu$ is the chemical potential, and $\hat{N}$ is the particle number operator. For evaluation of the shear viscosity it is sufficient to use the kinetic part of the stress tensor only
(see, e.g., Ref.~\cite{Enssetal}):
\begin{equation}
\hat{\Pi}_{\bm{q}=0}^{(xy)}=\sum_{\bm{p},\lambda=\uparrow,\downarrow}p_x p_y\hat a^\dagger_\lambda(\bm{p}) \, \hat a_\lambda^{}(\bm{p}).
\end{equation} 
The correlator $G_{\Pi}(\bm{q}=0,\tau)$ was evaluated for at least 61 points in imaginary time $\tau$, uniformly distributed in the interval $[0,\beta]$.
Since the correlator is symmetric with respect to $\beta/2$, we restrict the inversion 
procedure to the interval $\tau\in[0,\beta/2]$, and apply the symmetrization: 
$G_{\Pi}(\tau)\leftarrow[G_{\Pi}(\tau)+G_{\Pi}(\beta-\tau)]/2$ in order to decrease statistical fluctuations.

The inversion of \eqn{eqn:InversionProblem} represents numerically an ill-posed problem. However, by using known theoretical constraints the solution of this problem becomes easier to determine. Besides the non-negativity of the shear viscosity 
$\eta(\omega)\geqslant 0$, the sum rule
\begin{equation}
 \dfrac{1}{\pi}\int_{0}^{\infty}d\omega\left[ \eta(\omega) - \dfrac{C}{15\pi\sqrt{\omega}}\right] = \dfrac{\varepsilon}{3}, 
 \label{eqn:sumrule}
\end{equation} 
and the asymptotic tail behavior $\eta(\omega\to\infty)=C/(15\pi\sqrt{\omega})$ have been used as \apriori information. 
The Tan contact density $C$ and the energy density $\varepsilon$ are obtained consistently within the same simulation, 
and are in agreement with results reported in previous studies~\cite{BDM,Wlazlowskietal2}.
To perform the inversion we applied the methodology which combines two complementary methods:
singular value decomposition (SVD) and maximum entropy method (MEM), both described in Ref.~\cite{MagierskiWlazlowski}. The inversion procedure consists of two steps: (i) the SVD method provides us with the 
projection of the solution onto the subspace, where the inverse problem is well-posed $\tilde{\eta}$, (ii) the inversion 
is performed using the self-consistent MEM and we impose the SVD solution as additional external constraint, i.e., $P[\eta]=\tilde{\eta}$, where $P$ stands for the projection operator onto SVD subspace.
In order to estimate the stability of the method with respect 
to the algorithm parameters, the ``bootstrap" strategy was applied. For more details see the Supplemental Material of~\cite{Wlazlowskietal,Wlazlowskietal2}.

An important ingredient of the self-consistent MEM is an appropriately chosen class of \apriori models for the solution. 
We assume that the frequency dependent shear viscosity $\eta(\omega)$ has Lorentzian-like structure at low frequencies, smoothly evolving into the asymptotic tail behavior: 
\begin{eqnarray}
 M(\omega,\{\mu,\gamma,c,\alpha^{}_{1},\alpha^{}_{2}\}) = f(\omega,\{\alpha^{}_{1},\alpha^{}_{2}\})\,\dfrac{C}{15\pi\sqrt{\omega}}\nonumber\\
 +[1\!-\!f(\omega,\{\alpha^{}_{1},\alpha^{}_{2}\})]\, \mathcal{L}(\omega,\{\mu,\gamma,c\}),
\end{eqnarray} 
where
\begin{equation}
  f(\omega,\{\alpha^{}_{1},\alpha^{}_{2}\})=e^{-\alpha^{}_{1}\alpha^{}_{2}}\dfrac{e^{\alpha^{}_{1}\omega}-1}{1+e^{\alpha^{}_{1}(\omega-\alpha^{}_{2})}}
\end{equation} 
and
\begin{equation}
 \mathcal{L}(\omega,\{\mu,\gamma,c\})=c\,\dfrac{1}{\pi} \, \frac{\gamma}{(\omega-\mu)^2 + \gamma^{2}}.
 \label{eqn:Lorentzian}
\end{equation} 
The parameters $\{\mu,\gamma,c,\alpha^{}_{1},\alpha^{}_{2}\}$  describe admissible degrees of freedom of
the model and are adjusted automatically in the self-consistent manner. This choice of model space is reasonable as it is compatible with the low frequency behavior of the $\eta(\omega)$ predicted by hydrodynamic fluctuations formalism~\cite{ChafinSchafer,RomatschkeYoung} and with results obtained within Luttinger-Ward theory~\cite{Enssetal}. Finally, we assume that there is no sharp structure in the shear viscosity $\eta(\omega)$
in the low frequency limit. This assumption is of crucial importance, as it is well known, that sharp structures can be 
overlooked by numerical analytic continuation \cite{MagierskiWlazlowski}. Clearly, note that adding $\epsilon\delta(\omega)$ to the shear viscosity $\eta(\omega)$ corresponds to increasing the imaginary time correlator by a constant value $\Delta G_{\Pi}=\epsilon/\pi\beta$.

\section{Results}

\begin{figure}
\includegraphics[width=\pictsize\columnwidth]{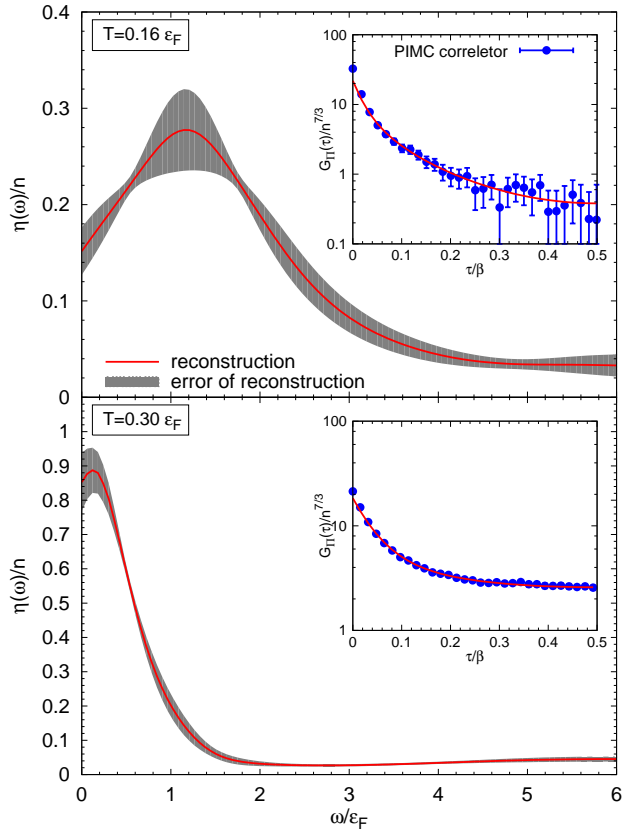}
\caption{ (Color online) The frequency dependent shear viscosity obtained from the QMC calculations for a $10^3$ lattice for two different temperatures: (upper panel) $T=0.16\,\eF$ (close to the critical temperature) and (lower panel) $T=0.30\,\eF$ (normal phase). The gray bands show uncertainty of the reconstruction. 
In the insets, the corresponding QMC correlators (blue points) are presented. The lines show the quality of the reproduction of the QMC data.
\label{fig:eta_dynamic} }
\end{figure}
In \fig{fig:eta_dynamic}, the dimensionless frequency dependent shear viscosity $\eta(\omega)/n$ is shown for two selected temperatures obtained for a $10^3$ lattice: close to the critical temperature of superfluid-to-normal phase transition 
$T_c=0.15(1)\,\eF$ and in the normal phase $T=0.30\,\eF$,  where 
$\eF=\frac{1}{2}k_F^2$ is the Fermi energy of the non-interacting gas.  In the normal phase $T\gtrsim 0.25\,\eF$, one may observe that $\eta(\omega)$ possesses a Lorentzian-like structure at low frequencies with a maximum in the vicinity 
of zero frequency. For temperatures below 
$T\lesssim 0.25\,\eF$ the suppression of $\eta(\omega)$ develops for the zero frequency. We associate this depletion with the onset of pairing correlations, which has been identified to exist in UFG above the critical temperature up to temperatures $T^*=0.20-0.25\,\eF$, commonly referred to as pseudogap regime~\cite{mag2009,mag2011,Wlazlowskietal2}. 

\begin{figure}
\includegraphics[width=\pictsize\columnwidth]{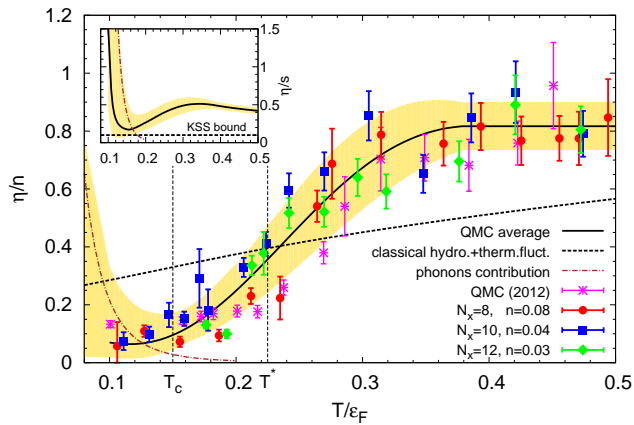}
\caption{ (Color online) The dimensionless static shear viscosity $\eta/n$ as a function of temperature for an $8^3$ lattice [solid (red) circles], a $10^3$ lattice [(blue) squares], and a $12^3$ lattice [(green) diamonds]. 
Vertical black dotted lines indicate the critical temperature of superfluid-to-normal phase transition for $T_c=0.15(1)\,\eF$ and the onset of Cooper pair formation, $T^*\approx0.22\,\eF$, respectively. The black line indicates smooth approximation of the QMC results and the (yellow) band reflects uncertainty of the viscosity computation. The (purple) asterisk shows the shear viscosity from the QMC calculations of Ref.~\cite{Wlazlowskietal}. 
The phonon contribution to the viscosity, evaluated according to Ref. \cite{RupakSchafer} is shown as a dot-dashed (brown) line and it is accounted for in the (yellow) band.
The hydrodynamic bound of Ref.~\cite{RomatschkeYoung} is plotted by the black dashed line.
The inset shows the ratio of the shear viscosity to entropy density ($\eta/s$) as a function of $T/\varepsilon^{}_{F}$, obtained from the smooth approximation of the QMC results. The black dashed line indicates KSS bound: $1/4\pi$.
\label{fig:viscosity} }
\end{figure}
In \fig{fig:viscosity}, the dimensionless static shear viscosity $\eta/n$ is shown
as a function of dimensionless temperature $T/\eF$. Results obtained for different lattices 
exhibit reasonable agreement, even though they are affected by significant uncertainty of the order of $10\%-20\%$.
For the $12^3$ lattice we provide results only for temperatures greater than $0.17\,\eF$ as we were not able to produce the correlator of required accuracy to perform reliable inversion for lower temperatures.
The temperature dependence of the shear viscosity can be resolved, which is clearly visible when a smooth approximation of QMC results is performed, while the amount of scattering of the QMC points can be used to estimate final uncertainty of the viscosity computation (yellow band in~\fig{fig:viscosity}). Moreover, the new results are consistent with previous estimate of the viscosity coefficient of Ref.~\cite{Wlazlowskietal}. We have checked the stability of our results with respect to the variation of the density, 
e.g., for the lattice size $10^3$ and $T \approx 0.16\,\eF$ and obtained 
$\eta/n\approx0.15(2),0.17(4),0.15(2)$ and $0.18(2)$ 
for densities $n\approx 0.04,0.06,0.08$ and $0.1$, respectively. In the low temperature regime no upturn originating from phonon contribution is observed, however, this does not imply that there is no upturn at all. Precisely, the boxes used in the simulations are not capable of properly capturing the phonon-related physics. It becomes clear if we note that the phonon occupation probability  $n_{\textrm{ph}}(k)\propto k^2/[\exp(ck/T)-1]$ (where $c$ is the sound velocity) has a peak at momenta of the order of $k \approx 2.2\,k_F\, T/\varepsilon_F$, which for $T/\varepsilon_F = 0.10\cdots 0.15$ is smaller than the smallest lattice momentum $2\pi/N_x$. Thus infrared momentum cutoff $2\pi/N_x$ removes almost entirely the pure phonon contribution. To take into account this source of the systematic error we modified the estimated error of the shear viscosity computation in such a way as to encapsulate the pure phonon contribution of Ref.~\cite{RupakSchafer}. Phonons appear to contribute significantly only well below the critical temperature $T_c$.  Above the critical temperature, when the system is in a normal state and the superfluid order parameter is vanishing the calculation approach adopted in Ref. \cite{RupakSchafer} is not valid anymore.  This approach is based on an effective theory for the phonon field alone, valid at very low temperatures only, where Cooper pair breaking and pseudogap physics can be ignored.  

\section{Comparison with the hydrodynamic bound}

Comparing our {\it ab initio} results with the hydrodynamic bound we find that as long as the system is in the normal phase (for $T>T^*$) the static shear viscosity is located above the bound. The results obtained for all three lattices indicate
that the hydrodynamic bound is violated when the system enters the pseudogap regime, which occurs above the critical temperature, at $T^*\approx 0.20-0.25\,\eF$. Interestingly, the corresponding ratio $\eta/s$ stays above the KSS bound $\eta/s\geqslant 1/4\pi\approx 0.08$ for all considered temperatures, with the minimal value $(\eta/s)^{}_{\textrm{min}}\approx 0.2$, located in the vicinity of the critical temperature. 
For the temperature evolution of the entropy we have used the MIT results of Ref.~\cite{MIT_exp}, as our results 
reproduce experimental data with reasonable accuracy (see~\cite{Wlazlowskietal,Drutetal}). 

The presented methodology, which produces the most probable results, together with the 
smoothness assumption of $\eta(\omega)$ clearly allows for the hydrodynamic bound to be violated in the low temperature regime 
for $T\lesssim T^*$. The hydrodynamics bound was obtained by adding the contribution due to thermal fluctuations to an undefined classical value of the viscosity \cite{ChafinSchafer,RomatschkeYoung} and with an amplitude depending on a cutoff parameter, which can be estimated only by order of magnitude, and thus a factor of order unity cannot be refuted. These estimates of a possible lower bound for viscosity are purely classical in character and they show that the role of classical thermal fluctuations could become important if the bare viscosity, evaluated in the absence of such thermal fluctuations, is small.  The bare viscosity can be evaluated in a classical kinetic approach, but to what extent such an estimate is valid in the case of a unitary Fermi gas remains an open question.  Even though a unitary Fermi gas is dilute in the sense that $\reff n^{1/3}\ll 1$, two-body collisions do not dominate
because $\sigma n^{2/3} ={\cal{O}}(1)$, where $\sigma \approx 4\pi/k_F^2$ is the interaction cross section between fermions at the Fermi surface. Except the role of very low momentum phonons at very low temperatures which cannot be properly included in QMC within the presently tractable lattice sizes, 
all other effects (pairing, pseudogap, fermion collisions) are included. As seen from Fig. \ref{fig:viscosity} phonons show a significant contribution only for temperatures below the critical temperature $T_c$. It implies that even if the  
phonons were included, the viscosity would have remained below the classical hydrodynamic bound \cite{ChafinSchafer,RomatschkeYoung} 
up to temperatures smaller than $T\approx 0.1\eF$. 
As another possibility, one can assume that hydrodynamic bound originates from long-wavelength
fluctuations, which are not included properly into our simulations due to finite size of the box. However, in this case we would expect to see a systematic trend in the results as we increase the box size, while we do not detect it.

\begin{figure}
\includegraphics[width=\pictsize\columnwidth]{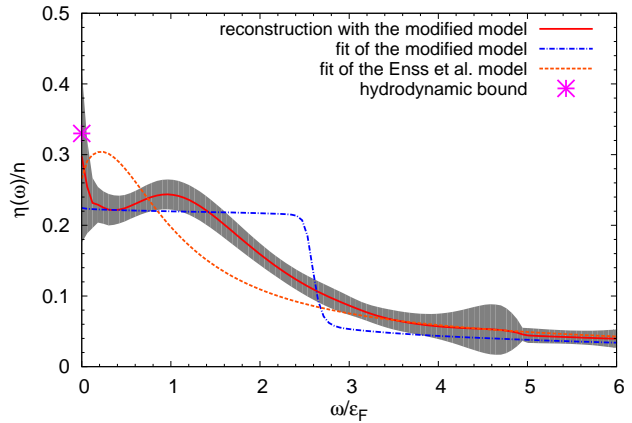}
\caption{ (Color online) The frequency dependent shear viscosity for the temperature $T=0.16\,\eF$ obtained by three different methodologies of the reconstruction as described in the text. The (purple) asterisk indicates the hydrodynamic bound obtained in Refs. \cite{ChafinSchafer,RomatschkeYoung}.
\label{fig:tests} }
\end{figure}

Even though we have no reason to expect that any alleged unaccounted for thermal fluctuations of the kind advocated in Refs. \cite{ChafinSchafer,RomatschkeYoung} might appear in the pseudogap regime below the temperature $T^*$, we included such a contribution into the spectral function to investigate whether our QMC data are consistent with this classical hydrodynamical bound. 
Three tests were performed using data obtained for a $10^3$ lattice 
at the critical temperature. Hydrodynamic fluctuations~\cite{ChafinSchafer,RomatschkeYoung}
suggest nonanalytic contributions for small frequencies of  the form
\begin{equation}
 \eta(\omega) = \eta(0) - \sqrt{\omega}\,T\, 
  \frac{7+\left(\frac{3}{2}\right)^{3/2}}{240\pi D_\eta^{3/2}},
\end{equation} 
where the momentum diffusion constant $D_\eta$ is positive. Thus, $\eta(\omega)$ is predicted to be a monotonically decreasing function for small frequencies. In the first test, the MEM solution is forced to be consistent with this assumption by replacing the Lorentzian-like structure at low frequencies~(\ref{eqn:Lorentzian}) by the expression
\begin{equation}
 \tilde{\mathcal{L}}(\omega,\{A,B\})=A-B\sqrt{\omega},
 \label{eqn:LorentzianM}
 \end{equation}
 where $A\geq 0, \; B\geq 0$.
This modification implies that the space for default models contains only solutions which are monotonically decreasing and explicitly consistent with predictions taking into account hydrodynamic fluctuations. In addition the constraint
provided by SVD is removed. The resulting frequency dependent viscosity is shown with a (red) solid  line in Fig. \ref{fig:tests}.
In the second test, the modified model is fit to the imaginary time correlator, in a similar way as has been done in Ref.~\cite{RomatschkeYoung}. This method typically violates the sum rule~(\ref{eqn:sumrule}) and the resulting $\eta(\omega)$ is shown with a (blue) dot-dashed line in~\fig{fig:tests}. It is  noted that our fit reveals a significantly different result than reported in Ref.~\cite{RomatschkeYoung}, which may be related to the fact that we use different models for the fitting procedure.
Finally, the QMC correlator was fit to the model proposed by Enss {\it et al.} in the paper~\cite{Enssetal}
\begin{equation}
 \eta(\omega)=\frac{W\tau_\eta}{1+(\omega\tau_\eta)^2} 
 + \frac{C}{15\pi\sqrt{\omega}} 
 \, \frac{\omega\tau_\eta(1+\omega\tau_\eta)} {1+(\omega\tau_\eta)^2},
 \label{eqn:EnssModel}
\end{equation} 
where the only free parameter is the viscous transport scattering time $\tau_\eta$, as the total Drude weight $W$ is fixed by the sum rule~(\ref{eqn:sumrule}) [see (red) dotted line in~\fig{fig:tests}].  All three approaches are able to reproduce the imaginary time correlator within its error bars. These simple tests result in a static shear viscosity below the hydrodynamic bound.
For MEM with the model (\ref{eqn:LorentzianM}), we find that $\eta/n$ is typically located below the bound.
However, we also find instances of the MEM algorithm that produce results above the bound as reflected in the error band depicted in~\fig{fig:tests}. Similar results were determined for other temperatures.

Based on these tests we conclude that the QMC results are compatible with the hydrodynamic bound only if we assume 
the existence of a very narrow peak located around $\omega\approx 0$.
The classical hydrodynamic fluctuation formalism predicts this type of peak, which has its 
origin in nonanalytic contributions to the shear viscosity. 
In Ref.~\cite{ChafinSchafer} it was estimated that the width of the nonanalytic structure can be of the order of $\Delta\omega \simeq 0.3 T(n/\eta)$ and at the critical temperature it gives $\Delta\omega/\eF\simeq 0.15$, which is in agreement with our first test.
However, we have to point out several important aspects of our results. 
(i)  We do not observe any abrupt change in the frequency dependent shear viscosity profile as we decrease the temperature, which is a typical signal of losing the resolution. Instead, the profile we construct evolves smoothly between the two cases presented in~\fig{fig:eta_dynamic}.
(ii) Our QMC simulations admit the existence of the pseudogap regime for the UFG~\cite{mag2009,mag2011,Wlazlowskietal2}. Other theoretical approaches
that admit the presence of the pseudogap predict the shear viscosity to be significantly suppressed, as in systems where a fermionic gap opens at $T^*$, and exhibit a clear violation of the hydrodynamic bound~\cite{BruunSmith,Guoetal}. Similar behavior was predicted for the spin-diffusion coefficient~\cite{Wulin}. Our {\it ab initio} results are qualitatively consistent with these observations and support scenarios provided by these theories. (iii) We find it difficult to rationalize the existence of a nonanalytical classical hydrodynamical correction to the spectral function deep in the quantum regime. 

\section{Conclusions}

In summary, we have presented the temperature evolution of a shear viscosity obtained within fully {\it ab initio} simulations. Assuming the smoothness of the $\eta(\omega)$ the results violate the hydrodynamic bound for $\eta/n$ in a low temperature regime, while the ratio $\eta/s$ stays above the KSS bound for all studied temperatures. Violation of the hydrodynamic bound occurs simultaneously with the occurrence of the Cooper pairs in the system. The accuracy of the QMC signal allows a result with the static viscosity being above the hydrodynamic bound only if we assume an existence of a sharp structure in the $\eta(\omega)$ located in the vicinity of zero frequency.

\section{Acknowledgments}

We thank P. Romatschke for making the results of Ref.~\cite{RomatschkeYoung} available to us. We acknowledge support under U.S. DOE Grants No. DE-FG02-97ER41014 and No. DE-FC02-07ER41457, and Contract No. N N202 128439 of the Polish Ministry of Science. One of the authors (G.W.) acknowledges the Polish Ministry of Science for support within the program ``Mobility Plus \!-\! I edition'' under Contract No. 628/MOB/2011/0. Calculations reported here have been performed at the Interdisciplinary Centre for Mathematical and Computational Modelling (ICM) at Warsaw University and on the University of Washington Hyak cluster funded by the NSF MRI Grant No. PHY-0922770. This research also used resources of the National Center 
for Computational Sciences at Oak Ridge National Laboratory, 
which is supported by the Office of Science of the Department 
of Energy under Contract No. DE-AC05-00OR22725.



\begin{thebibliography}{99}
\bibitem{stewart2010} 
	J.T. Stewart, J.P. Gaebler, and D.S. Jin, 
	Nature {\bf 454}, 744 (2010).
\bibitem{gaebler2010} 
	J. P. Gaebler, J. T. Stewart, T. E. Drake, D. S. Jin, A. Perali, P. Pieri, and G. C. Strinati, 
	Nature Phys. {\bf 6}, 569 (2010).
\bibitem{perali} 
	A. Perali, \etal 
	Phys. Rev. Lett. {\bf 106}, 060402 (2011).
\bibitem{Turlapov} 
	A. Turlapov, J. Kinast, B. Clancy, L. Luo, J. Joseph, J.E. Thomas, 
	J. Low Temp. Phys. {\bf 150}, 567 (2008).
\bibitem{Cao1} 
	C. Cao, E. Elliott, J. Joseph, H. Wu, J. Petricka, T. Schaefer, J. E. Thomas, 
	Science {\bf 331}, 58 (2011).
\bibitem{Cao2} 
	C. Cao, E. Elliott, H. Wu and J.E. Thomas, 
	New J. Phys. {\bf 13}, 075007 (2011).
\bibitem{reviews1} 
	S Giorgini, L.P. Pitaevskii, S. Stringari, 
	Rev. Mod. Phys.  {\bf 80}, 1215 (2008).
\bibitem{reviews2} 
        I. Bloch, J. Dalibard, W. Zwerger, 
        Rev. Mod. Phys. {\bf 80}, 885 (2008).
\bibitem{SchaferTeaney} 
	T. Sch\"{a}fer and D. Teaney, 
	Rep. Prog. Phys. {\bf 72}, 126001 (2009).
\bibitem{bcsbec} 
	\textit{The BCS-BEC crossover and the unitary Fermi Gas} 
	Lecture Notes in Physics, edited by W. Zwerger (Springer-Verlag, Berlin, 2012), Vol. 836.
\bibitem{FirstExperiments}
	\textit{Ultracold Fermi Gases}, Proceedings of the International School
	of Physics ``Enrico Fermi,'' CourseCLXIV, 
	Varenna, June 20-30,2006, edited by M. Inguscio, W. Ketterle, and C. Salomon
	(IOS Press, Amsterdam, 2008).
\bibitem{Carlsonetal}
	J. Carlson {\it et al.}, 
	Phys. Rev. A {\bf 84}, 061602(R) (2011).
\bibitem{VanHouckeetal} 
	K. Van Houcke, F. Werner, E. Kozik, N. Prokof'ev, B. Svistunov, M. J. H. Ku,
	A. T. Sommer, L. W. Cheuk, A. Schirotzek, and M. W. Zwierlein,
	Nature Phys. {\bf 8}, 366 (2012). 
\bibitem{Drutetal} 
	J.E. Drut, T.A. L\"{a}hde, G. Wlaz\l{}owski, P. Magierski, 
	Phys. Rev. A {\bf 85}, 051601(R) (2012).
\bibitem{CarlsonReddy} 
	J. Carlson and S. Reddy, 
	Phys. Rev. Lett. {\bf 95}, 060401 (2005).
\bibitem{GezerlisCarlson} 
	A. Gezerlis and J. Carlson, 
	Phys. Rev. C {\bf 77}, 032801(R) (2008).
\bibitem{mag2009}
	P. Magierski, G. Wlaz{\l}owski, A. Bulgac, and J.E. Drut, 
	Phys. Rev. Lett. {\bf 103}, 210403 (2009).
\bibitem{mag2011} 
	P. Magierski, G. Wlaz{\l}owski, and A. Bulgac, 
	Phys. Rev. Lett. {\bf 107}, 145304 (2011).
\bibitem{Burovskietal} 
	E. Burovski, N. Prokof'ev, B. Svistunov, and M. Troyer, 
	Phys. Rev. Lett. {\bf 96}, 160402 (2006).
\bibitem{Burovskietal2} 
	E. Burovski, E. Kozik, N. Prokof'ev, B. Svistunov, and M. Troyer, 
	Phys. Rev. Lett. {\bf 101}, 090402 (2008).
\bibitem{BDM} 
	A. Bulgac, J.E. Drut, and P. Magierski, 
	Phys. Rev. A {\bf 78} 023625 (2008).
\bibitem{Wlazlowskietal} 
	G. Wlaz{\l}owski, P. Magierski, and J.E. Drut, 
	Phys. Rev. Lett. {\bf 109}, 020406 (2012).
\bibitem{Wlazlowskietal2} 
	G. Wlaz{\l}owski, P. Magierski, J.E. Drut, A. Bulgac, and K.J. Roche,
	Phys. Rev. Lett. {\bf 110}, 090401 (2013).
\bibitem{KSS} 
	P.K. Kovtun, D.T. Son, and A.O. Starinets, 
	Phys. Rev. Lett. {\bf 94}, 111601 (2005).
\bibitem{BruunSmith} 
	G.M. Bruun and H. Smith, 
	Phys. Rev. A {\bf 75}, 043612 (2007).
\bibitem{Guoetal} 
	H. Guo, D. Wulin, C.-C. Chien, and K. Levin, 
	Phys. Rev. Lett. {\bf 107}, 020403 (2011); 
	New J. Phys. {\bf 13}, 075011 (2011).
\bibitem{RupakSchafer} 
	G. Rupak and T. Sch\"{a}fer, 
	Phys. Rev. A {\bf 76}, 053607 (2007).
\bibitem{ChafinSchafer} 
	C. Chafin, T. Sch\"{a}fer,
	Phys. Rev. A {\bf 87}, 023629 (2013).
\bibitem{RomatschkeYoung}
	P. Romatschke, R.E. Young,
	Phys. Rev. A {\bf 87}, 053606 (2013).   
\bibitem{Enssetal} 
	T. Enss, R. Haussmann, W. Zwerger, 
	Ann. Phys. {\bf 326}, 770 (2011).
\bibitem{MagierskiWlazlowski} 
	P. Magierski and G. Wlaz\l{}owski, 
	Comput. Phys. Commun. {\bf 183}, 2264 (2012).
\bibitem{MIT_exp}
	M.J.H.~Ku, A.T.~Sommer, L.W.~Cheuk, M.W.~Zwierlein,
	Science {\bf 335}, 563 (2012).
\bibitem{Wulin} 
	D. Wulin, H. Guo, C.-C. Chien, and K. Levin, 
	Phys. Rev. A {\bf 83}, 061601(R) (2011).
\end{thebibliography}
\end{document}